\documentclass[twocolumn,aps,floatfix,superscriptaddress,amsfonts,amssymb,superscriptaddress]{revtex4}
\usepackage{color}

\usepackage{graphicx}
\usepackage{amsmath}
\usepackage{latexsym}
\usepackage{epstopdf}
\usepackage{amsmath}
\usepackage{amssymb,amsmath}
\usepackage{multirow}
\usepackage{mathtools}
\newcommand{\beq}{\begin{equation}}
\newcommand{\eeq}{\end{equation}}

\usepackage{lmodern} \usepackage[T1]{fontenc}
\usepackage{tikz}
\usepackage{xcolor}
\usepackage{hyperref}
\usetikzlibrary{patterns}
\newcommand{\vect}[1]{\boldsymbol{#1}}
\newcommand{\tave}{t_{\mathrm{ave}}}
\newcommand{\trel}{t_{\mathrm{rel}}}
\newcommand{\tc}{t_{\mathrm{c}}}
\newcommand{\te}{t_{\mathrm{E}}}

\newcommand{\glocr}{g_{\mathrm{loc}}^{\mathrm{R}}}

\begin{document}

\title{Emergence of Floquet behavior for lattice fermions driven by light pulses}
\author{Mona H. Kalthoff}
\affiliation{Department of Physics, Georgetown University, 37th St. and O St., NW,  Washington, D.C. 20057, USA }
\affiliation{Lehrstuhl f\"ur Theoretische Physik, I, Technische Universit\"at Dortmund, Otto-Hahn Stra\ss{}e 4, 44221 Dortmund, Germany}

\author{G\"otz S. Uhrig}
\affiliation{Lehrstuhl f\"ur Theoretische Physik, I, Technische Universit\"at Dortmund, Otto-Hahn Stra\ss{}e 4, 44221 Dortmund, Germany}
\author{J. K. Freericks}
\affiliation{Department of Physics, Georgetown University, 37th St. and O St., NW,  Washington, D.C. 20057, USA }
\date{23 July 2018}

\begin{abstract}
As many-body Floquet theory becomes more popular, it is important to find ways to connect theory
with experiment. Theoretical calculations can have a periodic driving field that is always on, but experiment
cannot. Hence, we need to know how long a driving field is needed before the system starts to look
like the periodically driven Floquet system. We answer this question here for noninteracting
band electrons in the infinite-dimensional limit by studying the properties of the system under pulsed
driving fields and illustrating how they approach the Floquet limit. Our focus is on determining the minimal pulse lengths needed to recover the qualitative and semiquantitative Floquet theory results.
\\\quad\\
The final publication is available at Physical Review B via \url{https://link.aps.org/doi/10.1103/PhysRevB.98.035138} 
\end{abstract}
\maketitle

\section{Introduction}
Floquet theory has a long history, going back to the late 1800s \cite{Floquet1883}. Recently, it has become
a topic of wide interest in the condensed-matter community, especially with the relationship
between periodic driving and topological properties \cite{Lindner2011, Kitagawa2010}. Floquet systems require the driving
field to be present for all times. This presents a challenge experimentally, since the field must be turned
on and then off in realistic experiments. In addition, it is expected that interacting Floquet systems which have been turned on
for a long time will generally have runaway heating, and end up in the infinite-temperature limit.
This motivates the following question: How long does a pulsed field need to be in order to describe 
the Floquet regime well? We answer this question for noninteracting band electrons here. 

Experimentally, this is an important issue. Seminal work by the Gedik group showed how one
can transiently change the topology of a topological insulator when driven by circularly
polarized light \cite{Gedik2013}. Theory indicated how one can determine the band gaps that opened \cite{Fregoso2013}. 
But the theoretical premise of this work was that when we examine properties at the center of the pump pulse, they will look like the infinitely driven Floquet system. 
While this cannot be precisely true, it is approximately true. In this work, we examine this question in detail and determine
criteria for which one can approximate the Floquet regime well, and we also show how one can average transient results to recover Floquet behavior in cases where the Floquet limit does not
immediately emerge. 
We anticipate that much of these criteria will continue to hold when interactions are added, but provide no proof of that conjecture. 
Additionally we show that the noninteracting density of states of a periodically driven system is non-negative.

Some previous theory has examined these pulsed systems in the transient regime. One example is
a theoretical calculation in the change of the topology of graphene due to a circularly polarized electric field pulse \cite{Sentef2015} and another examined the transition metal dichalcogenides \cite{Claassen2016, Tang2017}. But none of that work addressed the specific question of how long must a pulse be on before the system appears to be described by the Floquet limit. We do so here.

We focus on examining band electrons driven by an external electric field. 
The problem is solved exactly via the Peierls' substitution \cite{Peierls1933}. We focus on the limit of infinite dimensions, because it
allows us to obtain a number of exact analytic relations. It also allows for this work to benchmark 
interacting calculations performed with nonequilibrium dynamical
mean-field theory \cite{Aoki2014, Freericks2006_3} in the future.

The remainder of the paper is organized as follows: In Sec. II, we introduce the model and the methodology used to solve for the retarded Green's functions for different pulsed drives. In Sec. III, we present our numerical results. Sec. IV has our conclusions. An appendix that proves nonnegativity of the time-averaged density of states follows at the end.

\section{Model}
We illustrate next how to describe lattice fermions under the influence of an external field. We start with the tight-binding Hamiltonian \cite{Slater1954} in the absence of a field given by 
\begin{equation}
\label{eqn:tight_binding_H}
\mathcal{H}_0=-\sum_{ij=1}^N t_{ij}c_i^\dagger c_j -\mu \sum_{i=1}^N c_i^\dagger c_i\,,
\end{equation}
where $t_{ij}$ is the Hermitian hopping matrix, $\mu$ is the chemical potential, and $N$ is the number of lattice sites.
The fermionic annihilation operator $c_j$ destroys an electron at lattice site $j$ while the fermionic creation operator $c_i^\dagger$ creates an electron at lattice site $i$. In this paper, we assume spinless electrons, and nearest-neighbor hopping on a hypercubic lattice in $d\rightarrow \infty$ dimensions.
 The nonzero elements $t_{ij}$ of the hopping matrix are given by 
 \begin{equation}
 t_{ij}=\frac{t^*}{2\sqrt{d}}
\end{equation}   
\cite{Metzner1989}, and depend on the rescaled hopping constant $t^*$. We couple this system to an external electromagnetic field described by
\begin{equation}
\vect{E}\left(\vect{r},t\right)=-\vect{\nabla}\Phi\left(\vect{r},t\right)-\frac{1}{c}\frac{\partial \vect{A}\left(\vect{r},t\right)}{\partial t}\,,
\end{equation}
where $\Phi\left(\vect{r},t\right)$ is a scalar potential and $\vect{A}\left(\vect{r},t\right)$ is a vector potential.
The speed of light is $c$, and we use the Hamiltonian gauge \cite{Jackson2001} to set the scalar potential $\Phi\left(\vect{r},t\right)=0$. The electric field effect on the hopping matrix is taken into account by performing the Peierls' substitution \cite{Jauho1984}. The original matrix element is multiplied by the exponential of the integral over the vector potential from the lattice vector $\vect{R}_i$ to the lattice vector $\vect{R}_j$ as follows:
\begin{equation}
t_{ij}\rightarrow t_{ij}\mathrm{exp}\left(-\frac{ie}{\hbar c}
\int_{\vect{R_i}}^{\vect{R}_j} \vect{A}\left(\vect{r},t\right)\,
\mathrm{d}\vect{r}\right)\,.
\end{equation}
Here, the absolute value of the electron charge is given by $e$. This Peierls' substitution is for a single band model, which means there are no dipole transitions between bands \cite{Turkowski2005}. While the electric fields we are considering vary in time, we assume they are spatially uniform, so the magnetic field associated with them is negligible and $\vect{A}\left(\vect{r},t\right)\rightarrow \vect{A}\left(t\right)$. 
This assumption can be made because the wavelength of the driving field is much larger than the atomic scales. 
In this case, the momentum-space representation for the Hamiltonian of noninteracting electrons coupled to a spatially uniform electric field can be written as a function of the band structure 
\begin{equation}
\varepsilon\left(\vect{k}\right)=
-\frac{t^*}{\sqrt{d}}\sum_{\alpha=1}^d \cos\left(k_\alpha a\right)\,,
\end{equation}
where $a$ is the lattice constant. The momentum-space Hamiltonian becomes
\begin{equation}
\label{eqn:Hamiltonian_k}
\mathcal{H}\left(t\right)=\sum_{\vect{k}} 
\left[\varepsilon
\left(\vect{k}-\frac{e\vect{A}\left(t\right)}{\hbar c}\right)-\mu\right]
c_{\vect{k}}^\dagger c_{\vect{k}}\,,
\end{equation}
with 
\begin{equation}
c_{\vect{k}}^\dagger=\frac{1}{\sqrt{N}}\sum_{n=1}^N 
\mathrm{exp}\left[-i\vect{R}_n\vect{k}\right]c_n^\dagger
\end{equation}
and the Hermitian conjugate equation for $c_{\vect{k}}$. 
For many driving fields, such as an electric field that is periodic in time, 
the Hamiltonian in Eq.~\eqref{eqn:Hamiltonian_k} is a Floquet Hamiltonian. 
It has periodic time dependence due to the time dependence it inherits from the electric field.
However, because the Hamiltonian with the Peierls' substitution is diagonal in momentum space, it commutes with itself for all times $\left[\mathcal{H}\left(t\right),\mathcal{H}\left(t'\right)\right]=0$. This greatly simplifies the problem. We consider the momentum-space representation of the creation and the annihilation operator in the Heisenberg picture, where $c_{\vect{k}}\left(t\right)=\mathrm{exp}\left[it\mathcal{H}\left( t\right)\right]c_{\vect{k}}\mathrm{exp}\left[-it\mathcal{H}\left( t\right)\right]$, and use the Hamiltonian in Eq. \eqref{eqn:Hamiltonian_k} to derive their time evolution, yielding
\begin{equation}
\label{eqn:creation_operator_timeevolution}
c_{\vect{k}}\left(t\right)=\mathrm{exp}
\left[
-\frac{i}{\hbar}
\int_{-\infty}^t
\left[\varepsilon
\left(\vect{k}-\frac{e\vect{A}\left(t\right)}{\hbar c}\right)-\mu\right]
\,\mathrm{d}t
\right]
c_{\vect{k}}\,.
\end{equation}
This result allows us to analytically calculate the retarded momentum-dependent Green's function, which is defined by 
\begin{equation}
\label{eqn:Def.Greensfunction}
g^\mathrm{R}\left(\vect{k},t_1,t_2\right)=
-\frac{i}{\hbar}\Theta\left(t_1-t_2\right)
\left\langle
\left\lbrace c_{\vect{k}}\left(t_1\right),c_{\vect{k}}^\dagger\left(t_2\right)
\right\rbrace_+
\right\rangle\,.
\end{equation}
The angular brackets denote thermal averaging $\left\langle O\right\rangle= \mathrm{Tr}\left[\mathrm{exp}\left(-\beta \mathcal{H}_0\right)O\right]/\mathrm{Tr}\left[\mathrm{exp}\left(-\beta \mathcal{H}_0\right)\right]$, where the inverse temperature is given by $\beta=1/T$ and $\mathcal{H}_0$ is the field-free Hamiltonian in Eq. \eqref{eqn:tight_binding_H}.
To calculate the momentum-dependent Green's function in Eq. \eqref{eqn:Def.Greensfunction}, we specialize to a vector potential that lies along the diagonal, introducing a scalar function $A\left(t\right)$ that is associated with the vector potential via $\vect{A}\left(t\right)=A\left(t\right)\left(1,1,1,\dots\right)$. In this case, the retarded momentum-dependent Green's function is given by
\begin{eqnarray}
\label{eqn:momentum_dependent_Greensfunction}
g^\mathrm{R}\left(\vect{k},t_1,t_2\right)&=&
-\frac{i}{\hbar}\Theta\left(t_1-t_2\right)e^{\frac{i\mu}{\hbar}\left(t_1-t_2\right)}
\\
&&\times \mathrm{exp}\left[-i\frac{\varepsilon\left(\vect{k}\right)}{\hbar}
\int_{t_2}^{t_1}\cos\left(\frac{eaA\left(t\right)}{\hbar c}\right)\,
\mathrm{d}t\right]\nonumber
\\
&&\times \mathrm{exp}\left[-i\frac{\tilde{\varepsilon}\left(\vect{k}\right)}{\hbar}
\int_{t_2}^{t_1}\sin\left(\frac{eaA\left(t\right)}{\hbar c}\right)\,
\mathrm{d}t\right]\,,\nonumber
\end{eqnarray}
using the complementary energy function 
\begin{equation}
\tilde{\varepsilon}\left(\vect{k}\right)=-\frac{t^*}{\sqrt{d}}\sum_{\alpha=1}^d
\sin\left(k_\alpha a\right)\,,
\end{equation}
which is the projection of the band velocity along the field direction.
Of course this retarded Greens function is independent of temperature as expected for Green's functions of noninteracting systems.
Note that in equilibrium the Hamiltonian is constant in time and hence the whole problem is time-translation invariant. Thus, only time differences matter and the Green's function depends solely on the relative time $t_{\mathrm{rel}}=t_1-t_2$. However, due to the coupling of the lattice fermions to a time-dependent electric field, the Green's function in Eq. \eqref{eqn:momentum_dependent_Greensfunction} depends separately on the time $t_2$ of the creation operator and the time $t_1$ of the annihilation operator. The local Green's function can be computed by summing over all momentum vectors $\vect{k}$, which corresponds to the integration over $\varepsilon$ and $\tilde{\varepsilon}$ respectively, by using the joint density of states for tight binding electrons on a hypercubic lattice
\begin{equation}
\rho_0\left(\varepsilon,\tilde{\varepsilon}\right)=
\left(\frac{1}{\sqrt{\pi} t^{*} a^{d}}\right)^2
\mathrm{exp}\left[-\left(\frac{\varepsilon}{t^{*}}\right)^2
-
\left(\frac{\tilde{\varepsilon}}{t^{*}}\right)^2
\right]\,.
\end{equation}
Lengthy calculations given in Ref. \cite{Turkowski2005} yield
\begin{equation}
\label{eqn:Greensfunction_with_I}
g_{\mathrm{loc}}^\mathrm{R}\left(t_1,t_2\right)=
-\frac{i}{\hbar}\Theta\left(t_1-t_2\right)e^{\frac{i\mu}{\hbar}\left(t_1-t_2\right)}
e^{-\left(\frac{t^*}{2\hbar}\right)^2\left|I\left(t_1,t_2\right)\right|^2},
\end{equation}
where $I\left(t_1,t_2\right)$ is given by 
\begin{equation}
\label{eqn:I_definition}
I\left(t_1,t_2\right)=\int_{t_2}^{t_1}\mathrm{exp}\left[\frac{ieaA\left(t\right)}{\hbar c}\right]\,\mathrm{d}t\,.
\end{equation}
In this paper, we will assume half filling ($\mu=0$), so the time-dependent local Green's function in Eq. \eqref{eqn:Greensfunction_with_I} is purely imaginary. 
The density of states (DOS) is found from the temporal Fourier transform of the local retarded Green's function. 
In equilibrium, where the Green's function is only dependent on the relative time $t_1-t_2$, it is unambiguous what is meant by the frequency-dependent response which is obtained by computing the Fourier transform of 
$\glocr\left(t_1-t_2\right)$.
If, however, a driving that varies in time is applied to the system, the situation is different, as we have a two-time response $\glocr\left(t_1,t_2\right)$.

In generic pump-probe experiments, frequency resolved quantities are measured as a function of the delay time.
Careful analysis of a given experiment will yield the proper way to integrate over time and construct the frequency-dependent response, as was done for photoemission in Ref. \cite{Freericks2016_2,Freericks2009_1,Gruber2001}.
Nevertheless, when we examine Green's functions, it is useful to represent them in terms of frequency irrespective of any particular measurement. This procedure is not unique, and we describe two particular ways to do it next.

\begin{figure}
    \includegraphics[width= \columnwidth]{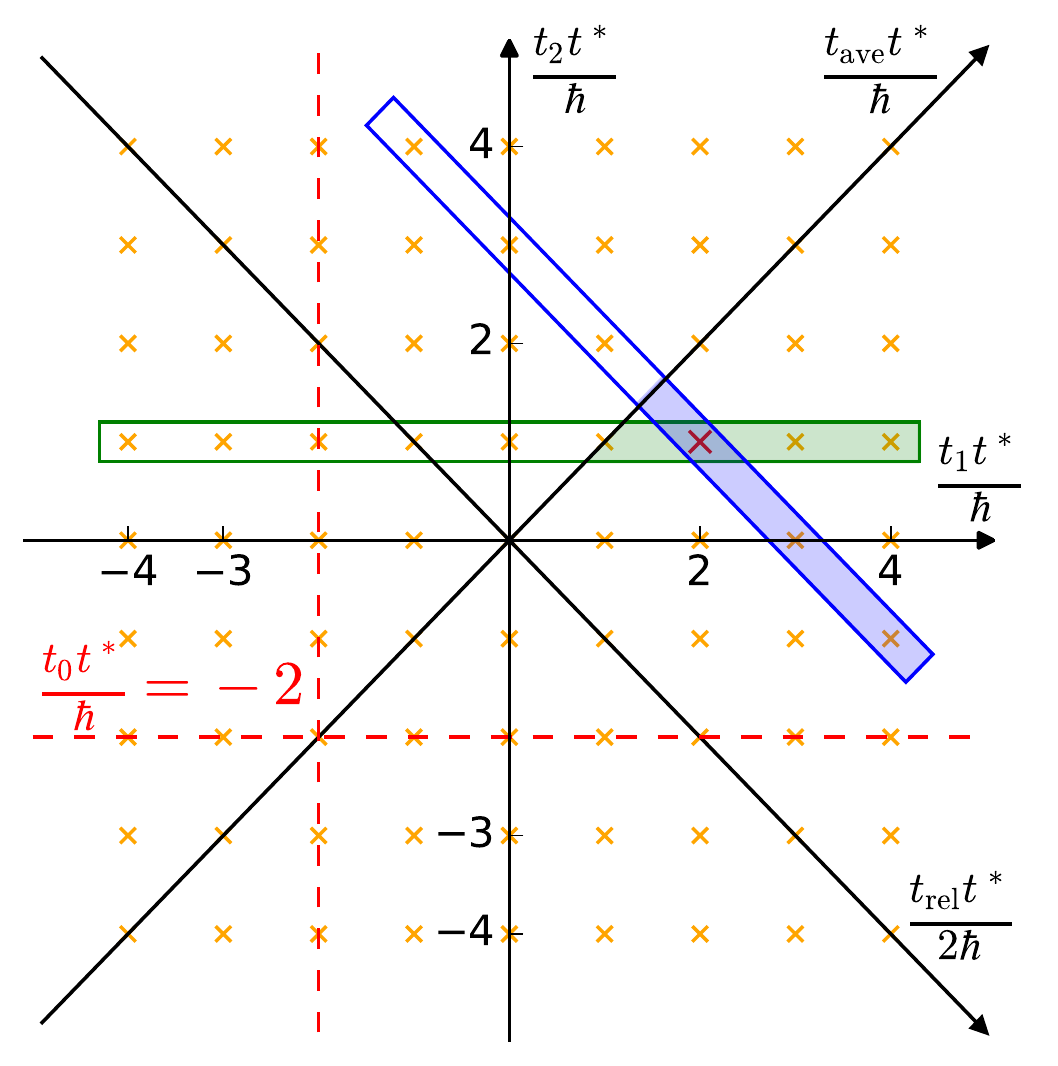}    
    \caption{(color online) Schematic display of the two times for the Green's function and the integration directions for the horizontal time and average time DOS. For retarded quantities, the Fourier transform runs only over the darker parts of the rectangles. Retarded quantities are nonzero only below and to the right of the diagonal labeled $\tave$, defined by $\trel=0$.}
    \label{fig:diff_fourier_transformations_model}
\end{figure}
The first definition introduces the Wigner coordinates \cite{Wigner1932}, where the Fourier transform is performed in the relative time $t_{\mathrm{rel}}=t_1-t_2$ while the average time $t_{\mathrm{ave}}=\left(t_2+t_1\right)/2$ is kept constant. The Green's function is expressed as
\begin{equation}
\glocr\left(t_1,t_2\right)=
\glocr\left(\tave+\frac{\trel}{2},\tave-\frac{\trel}{2}\right)\,.
\label{eqn:Green_diag}
\end{equation}
Figure \ref{fig:diff_fourier_transformations_model} schematically displays the concept of these coordinates by introducing the diagonal axis for $\tave$ and $\trel$. 
This means that all grid points on a line perpendicular to the diagonal axis for $\tave$ are associated with the same average time, just as all grid points perpendicular to the axis $\trel/2$ have the same relative time.
For example, the grid point $\left(t_1,t_2\right)=\left(2,1\right)\hbar/t^*$, which is marked in red, has the average time $\tave=1.5\hbar/t^*$, and so do all the grid points in the blue rectangle. That is, it is exactly those times that the Fourier transformation is computed over when the average time is chosen to be $\tave=1.5\hbar/t^*$, which is why we will refer to this as the diagonal Fourier transform $\mathcal{F}_\mathrm{D}$. It is employed to calculate the diagonal DOS via
\begin{equation}
\rho_\mathrm{D}\left(\omega,\tave\right)=-\frac{1}{\pi}\mathrm{Im}\left[
\int_0^\infty e^{i\omega t_{\mathrm{rel}}}
g_{\mathrm{loc}}^{\mathrm{R}}\left(t_{\mathrm{ave}},t_{\mathrm{tel}}\right)\,
\mathrm{d}t_{\mathrm{rel}}
\right]\,.
\label{eqn:DOS_diagonal}
\end{equation}
The above procedure is popular because $\tave$ can loosely be interpreted as the "time corresponding to the DOS"
In this case it is reasonable to identify the time associated with the Fourier transform to be in the middle of the interval $\left[t_2,t_1\right]$. 

There are potential problems with this choice. If the pulse starts at $t_0$, then for $\tave<t_0$ there are traces of the effect of the pulse in the DOS even though $\tave$ is before the pulse was turned on. 
This is due to large enough positive $\trel$ contributions in the Green's function given in Eq. \eqref{eqn:Green_diag} from times after the onset of the pulse $\left(t_1>t_0\right)$.
The converse is also true. If $\tave>t_0$, then for large enough $\trel$, we have $t_2<t_0$, so contributions to a field dressed DOS include terms before the field was turned on.

In Fig. \ref{fig:diff_fourier_transformations_model}, the dashed red line represents a pulse starting at $t_0=-2\hbar/t^*$. Even if the average time is chosen to be $\tave=1.5\hbar/t^*$, the Fourier transformation with respect to $\trel>0$, displayed as the area shaded in blue, will eventually cross the dashed line and include values $t_2<t_0$.

We can also define a horizontal Green's function as 
\begin{equation}
\glocr\left(t_1,t_2\right)=
\glocr\left(\trel+t_2,t_2\right)\,,
\label{eqn:Green_vert}
\end{equation}
and again perform the Fourier transform in the relative time $\trel$. In Figure \ref{fig:diff_fourier_transformations_model}, this is displayed as the green box, for all of the grid points in it satisfy $t_2=\hbar/t^*$. This horizontal Fourier transform $\mathcal{F}_\mathrm{H}$ yields the horizontal density of states, given by 
\begin{equation}
\rho_\mathrm{H}\left(\omega,t_2\right)=-\frac{1}{\pi}\mathrm{Im}\left[
\int_0^\infty e^{i\omega t_{\mathrm{rel}}}
\glocr\left(\trel+t_2,t_2\right)
\,
\mathrm{d}t_{\mathrm{rel}}
\right]\,.
\label{eqn:DOS_horizontal}
\end{equation}
The advantage of this definition is, that for $t_2>t_0$, all times used for the Fourier transform occur after the onset of the pulse. 
That is, for $\trel>0$ the shaded green area in Fig. \ref{fig:diff_fourier_transformations_model} will never cross the dashed red line.
The disadvantage is that the average time is not fixed.
Of course, in static equilibrium both response functions are equal and indistinguishable. 

\section{Results}
Floquet theory is applicable for quantum systems with a Hamiltonian that is invariant under time translations $t\rightarrow t+t_{\mathrm{period}}$, i.e. a Hamiltonian being a periodic function in time with the period $t_{\mathrm{period}}$ \cite{Grifoni1998}. It is based on the Floquet formalism \cite{Floquet1883} and is commonly used to study strongly driven periodic quantum systems. The Hamiltionian in Eq. \eqref{eqn:Hamiltonian_k} fulfills these conditions if and only if the driving field is strictly periodic in time, meaning it is turned on at $tt^*/\hbar=-\infty$ and stays on.
It is obvious that such a driving can never be realized in an experiment. Therefore we will start this section by introducing the properties of the DOS of lattice fermions coupled to an infinite sinusoidal driving, and compare those results to three field pumps that are not strictly periodic, but are experimentally feasible. 
The raw data for all figures can be found in the supplemental materials files~\cite{supp_material}.
\subsection{Infinite sinusoidal driving}
\label{subsec:infinite_driving}
\begin{figure}
    \includegraphics[width= \columnwidth]{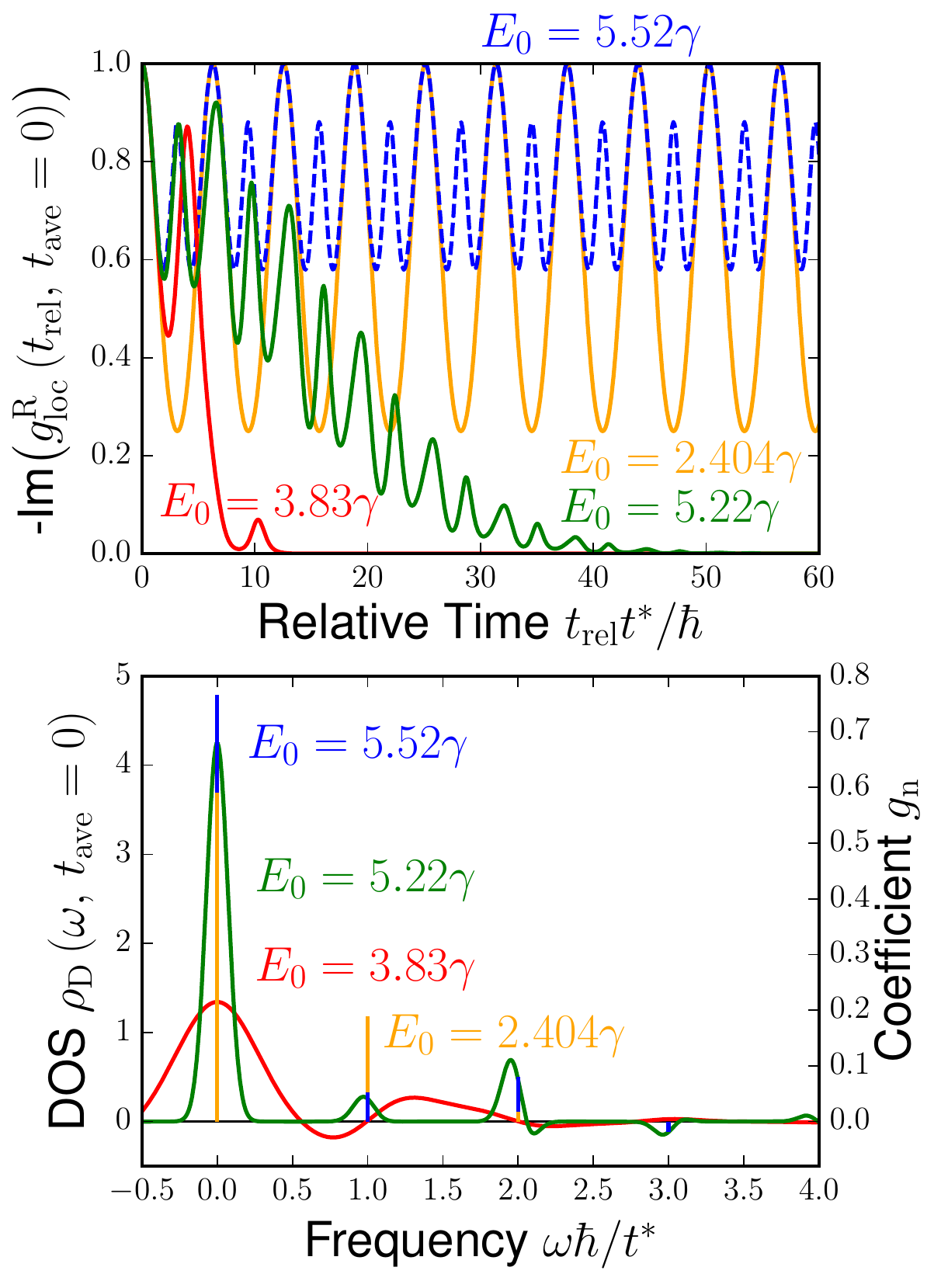}   
    \caption{
    (color online)
Upper panel: Negative imaginary part of the local retarded Green's function as a function of the relative time $\trel$ in units of the inverse rescaled hopping $\hbar /t^*$. 
Note that the real part vanishes because $\mu=0$. Lower panel: Diagonal DOS as a function of the frequency $\omega$ in units of the rescaled hopping. 
Because the DOS is a set of delta functions at $J_0\left(E_0/\gamma\right)=0$ the coefficients 
$g_{\mathrm{n}}=-1/\left(2\pi\right)\int_0^{2\pi}\mathrm{d}\trel\mathrm{Im}\left[ \glocr \left(\trel,0\right) \right]\cos\left(n\trel\right)$ 
of the Fourier series are displayed for $E_0=2.404\gamma$ and $E_0=5.52\gamma$. Other parameters: average time $\tave=0$, driving frequency $\gamma=t^*/\hbar$.}  
    \label{fig:infinite_sinusoidal}
\end{figure}
For the infinite sinusoidal driving with the frequency $\gamma$ and the amplitude $E$, the vector potential is given by 
\begin{equation}
\label{eqn:inf_vector_potential}
A_{\infty}=\frac{cE}{\gamma}\cos\left(\gamma t\right)\,,
\end{equation} 
whose simple form enables the analytic determination of the local Green's function.
If we define the modified amplitude $E_0=eaE/\hbar$, the squared absolute value of $I\left(t_1,t_2\right)$ in Eq. \eqref{eqn:I_definition} yields
\begin{eqnarray}
\label{eqn:I_infinite_sin_driving_withoutbessel}
\left|I_\infty\left(t_1,t_2\right)\right|^2&=&
\frac{1}{\gamma^2}\left|\int_{t_2\gamma}^{t_1\gamma}
f_c\left(t\right)\,\mathrm{d}t\right|^2 
\\\nonumber
&&+
\frac{1}{\gamma^2}\left|\int_{t_2\gamma}^{t_1\gamma}
f_s\left(t\right)\,\mathrm{d}t\right|^2\,.
\end{eqnarray}
with the integrands 
\begin{subequations}
\label{eqn:integrands_f}
\begin{eqnarray}
f_c\left(t\right)&=&\cos\left(
\frac{E_0}{\gamma}\cos\left(t\right)\right)
\\
f_s\left(t\right)&=&\sin\left(
\frac{E_0}{\gamma}\cos\left(t\right)\right)\,.
\end{eqnarray}
\end{subequations}
Both $f_c\left(t\right)$ and $f_s\left(t\right)$ are even functions that are $2\pi$-periodic, therefore they can be expressed as Fourier series with the Fourier coefficients $c_m$ and $s_m$ according to
\begin{subequations}
\begin{eqnarray}
f_c\left(t\right)&=&\frac{c_0}{2}+\sum_{m=1}^\infty c_m\cos\left(mt\right)
\\
f_s\left(t\right)&=&\sum_{m=1}^\infty s_m\cos\left(mt\right)\,.
\end{eqnarray}
\end{subequations}
Note that due to the fact that $f_s\left(t\right)$ is not only $2\pi$-periodic, but also $\pi$-anti-periodic, the coefficient $s_0$ vanishes, while the $\pi$-periodic function $f_c\left(t\right)$ has the coefficient $c_0=2J_0\left(E_0/\gamma\right)$, where $J_\alpha$ is the Bessel function of the first kind. Using the two $2\pi/\gamma$-periodic functions derived from integrating the Fourier coefficients of $f_c$ and $f_s$,
\begin{subequations}
\label{eqn:phi_def}
\begin{eqnarray}
\phi_c\left(t\right)&=&\frac{1}{\gamma}\sum_{m=1}^\infty \frac{c_m}{m}\sin\left(m\gamma t\right)
\\
\phi_s\left(t\right)&=&\frac{1}{\gamma}\sum_{m=1}^\infty \frac{s_m}{m}\sin\left(m\gamma t\right)\,,
\end{eqnarray}
\end{subequations}
the integration in Eq. \eqref{eqn:I_infinite_sin_driving_withoutbessel}
can now easily be computed, yielding
\begin{eqnarray}
\label{eqn:I_infinite_sin_driving_withbessel}
\left|I_\infty\left(t_1,t_2\right)\right|^2&=&
\left|\phi_s\left(t_1\right)-\phi_s\left(t_2\right)\right|^2\\
&+&\nonumber
\left|J_0\left(\frac{E_0}{\gamma}\right)\trel+\phi_c\left(t_1\right)-\phi_c\left(t_2\right)\right|^2.
\end{eqnarray}
In this form, it is obvious that the increase in $\left|I_\infty\left(t_1,t_2\right)\right|^2$ for large relative times is solely caused by the term dependent on the Bessel function and proportional to $\trel^2$;
a large $\left|I_\infty\left(t_1,t_2\right)\right|^2$ corresponds to a small Green's function as seen in Eq. \eqref{eqn:Greensfunction_with_I}.
This is because the periodic functions $\phi$ merely oscillate in time. Hence, if the amplitude and the frequency of the driving field are chosen in such a way that the Bessel function $J_0\left(E_0/\gamma\right)$ is zero, 
then the dephasing of the Green's function, corresponding to the decay of $g_{\mathrm{loc}}^\mathrm{R}\left(t_1,t_2\right)$ for large $\trel$, no longer occurs.

Figure \ref{fig:infinite_sinusoidal} shows the imaginary part of the time-dependent local Green's function at $\tave=0$, as defined in Eq. \eqref{eqn:Green_diag} for different amplitudes of the electric field, while the frequency is kept constant at $\gamma=t^*/\hbar$. 
Note that the Green's function at half filling is purely imaginary [see Eq. \eqref{eqn:Greensfunction_with_I}].
In this case, the absolute value of Bessel function is purely dependent on the amplitude of the driving. 
If the amplitude is chosen to be $E_0=2.404\gamma$, which corresponds to the first zero of the Bessel function (displayed in orange), 
it is clear that there is no dephasing in $\trel$ and the local Green's function oscillates with a period of $2\pi/\gamma$ between one and a constant value less than one.
Note that if the Bessel function is zero, the squared absolute value of $I_\infty\left(t_1,t_2\right)$ in Eq. \eqref{eqn:I_infinite_sin_driving_withbessel} is merely a superposition of sinusoidal functions that are periodic in $2\pi/\gamma$.
This means the DOS is a set of delta functions and the dominant frequencies are $\omega=0$ and $\omega=\pm \gamma$. 
The next zero of the Bessel function occurs at $E_0=5.52\gamma$, 
and again the imaginary part oscillates with a period of $2\pi/\gamma$ around a constant value. However, the amplitude changes and additional peaks appear. Therefore the DOS is again a set of delta functions, but it consists of more delta peaks than at the first zero. This behavior continues, as the amplitude takes values of higher zeros of the Bessel function.

The imaginary part of the Green's function at a constant $t_2=0$ [as defined in Eq. \eqref{eqn:Green_vert}] shows the same overall properties when displayed as a function of $\trel$, differences being that the lower extreme value shifts and the oscillations have double the frequency. 
This behavior can easily be understood, because in Eq. \eqref{eqn:I_infinite_sin_driving_withbessel} $t_1$ and $t_2$ are the arguments of the function $\phi$, where they solely appear in the argument of the sine function. But in the average time Green's function \eqref{eqn:Green_diag}, both $t_1$ and $t_2$ are dependent on $\trel/2$, while they are dependent on $\trel$ without an additional factor in the Green's function for the horizontal case \eqref{eqn:Green_vert}. 
Therefore the period of the oscillations in $\trel$ for a constant $\tave$ is twice as large for a constant $t_2$.

Contrary to the behavior at the zeros of the Bessel function, the dephasing in $\trel$ is fast at extreme values of the Bessel function like its first minimum at $E_0=3.83\gamma$, which is displayed in red, because the squared absolute value of $\left|I_\infty\left(t_1,t_2\right)\right|^2$ is large, even if $\trel$ is still comparatively small. As the argument of the Bessel function becomes smaller, the dephasing takes longer. 

For a time-independent Hamiltonian it is easy to prove that the DOS is positive semidefinite, via the Lehmann representation. However, this is not necessarily the case for the DOS of a driven system where the DOS takes negative values if it is computed at a constant $\tave$ or $t_2$. 
On the other hand, for a pure Floquet Hamiltonian the DOS has to be periodic in the Floquet period, which is the period of the driving, and averaging over this Floquet period in $\tave$ or $t_2$, respectively, does lead to a semidefinite DOS. 
To show this analytically, we consider the retarded Green's function at half filling, introduced in Eq. \eqref{eqn:momentum_dependent_Greensfunction} and write it in terms of the functions $\phi$ introduced in Eq. \eqref{eqn:phi_def}, yielding
\begin{eqnarray}
g^\mathrm{R}\left(\vect{k},t_1,t_2\right)&=&
-\frac{i}{\hbar}\Theta\left(t_1-t_2\right)
\\
&&\times 
\nonumber
\mathrm{exp}\left[-\frac{i}{\hbar}
\varepsilon\left(\vect{k}\right) J_0\left(\frac{E_0}{\hbar}\right)\left(t_1-t_2\right)
\right]
\\
&&\times 
\nonumber
\mathrm{exp}\left[-\frac{i}{\hbar}
\left(
\varepsilon\left(\vect{k}\right) \phi_c\left(t_1\right)
-
\varepsilon\left(\vect{k}\right) \phi_c\left(t_2\right)\right)\right]
\\
&&\times 
\mathrm{exp}\left[-\frac{i}{\hbar}
\left(
\tilde{\varepsilon}\left(\vect{k}\right)  \phi_s\left(t_1\right)
-
\tilde{\varepsilon}\left(\vect{k}\right)  \phi_s\left(t_2\right)
\right)\right]
\,.\nonumber
\end{eqnarray}  
Defining the $2\pi/\gamma$ periodic function $\Phi\left(t,\vect{k}\right)$
\begin{subequations} 
\begin{eqnarray}
\Phi\left(t,\vect{k}\right)
&=&
\mathrm{exp}
\left[
-\frac{i\varepsilon\left(\vect{k}\right)}{\hbar} \phi_c\left(t\right)
-\frac{i\tilde{\varepsilon}\left(\vect{k}\right)}{\hbar} \phi_s\left(t\right)
\right]
\\
&=&
\sum_m
e^{im\gamma t}
f_m\left(\vect{k}\right)
\end{eqnarray}
\end{subequations}
allows us to write this Green's function as 
\begin{eqnarray}
\label{eqn:Greens_proof}
g^\mathrm{R}\left(\vect{k},\tave,\trel\right)
=\nonumber
-\frac{i}{\hbar}
\mathrm{exp}\left[
-\frac{i\varepsilon\left(\vect{k}\right)}{\hbar} J_0\left(\frac{E_0}{\hbar}\right)\trel\right]
\\
\times \Theta\left(\trel\right)
\Phi^*\left(\tave-\frac{\trel}{2},\vect{k}\right)\Phi\left(\tave+\frac{\trel}{2},\vect{k}\right)\,.
\end{eqnarray}
As shown in the Appendix, using the convolution of two $2\pi/\gamma$ periodic functions and the fact that $\Phi$ can be written as a Fourier series with the coefficients $f_m$,
the integral over one period is given by 
\begin{equation}
\frac{\gamma}{2\pi}\int_x^{x+\frac{2\pi}{\gamma}}
\Phi^*\left(\tilde{t}-\frac{t}{2}\right)\Phi\left(\tilde{t}+\frac{t}{2}\right)\mathrm{d}\tilde{t}
=
\sum_{m}
|f_m|^2 e^{imt\gamma}\,.
\end{equation}
Note that we are suppressing the $\vect{k}$ dependence of both $\Phi$ and $f_m$ in order to simplify the notation. This allows us to compute the averaged Green's function 
\begin{equation}
\label{eqn:averaged_Greensfunction}
\bar{g}^R\left(\vect{k},\trel\right)=\frac{\gamma}{2\pi}\int_x^{x+\frac{2\pi}{\gamma}}g^R\left(\vect{k},\tau,\trel\right)\mathrm{d}\tau
\end{equation} 
and the averaged spectral function $\bar{A}\left(\omega,\vect{k}\right)$, yielding
\begin{subequations}
\begin{eqnarray}
\label{eqn:averaged_Spectralfunction}
\bar{A}\left(\omega,\vect{k}\right)=
-\frac{1}{\pi}\mathrm{Im}\left(
\int_0^\infty 
e^{i\omega \trel} \bar{g}^R\left(\vect{k},\trel\right)
\mathrm{d}\trel
\right)\\
=
\frac{1}{\hbar}
\sum_{m}
|f_m|^2
\delta\left(
\omega+m\gamma-\frac{\varepsilon\left(\vect{k}\right)}{\hbar}
J_0\left(\frac{E_0}{\hbar}\right)
\right)\,
\end{eqnarray}
\end{subequations}
which is indeed non-negative for all $\omega$.
While the diagonal and the horizontal DOS corresponding to the infinite sinusoidal driving are different at a given time $t_2$ for the horizontal DOS and $\tave$ for the diagonal DOS (even if $t_2=\tave$), the time-averaged spectral function $\bar{A}\left(\omega,\vect{k}\right)$ (and therefore the DOS averaged over the Floquet period) are always the same. 
Details can be found in the Appendix.
\subsection{Semi-infinite sinusoidal driving starting at $t_0=0$}
\label{subsec:semi_infinite_driving}
While a driving field that is switched on at a given time $t_0=0$ but stays on is also not experimentally implementable, it is useful to study the properties of its DOS because there are many similarities to the behavior of the DOS of driving pulses that can be experimentally implemented (see below).
The vector potential of this semi-infinite sinusoidal driving is given by $A=\left(cE/\gamma\right)\cos\left(\gamma t\right)\Theta\left(t\right)-\left(cE/\gamma\right)\Theta\left(t\right)$. Again, the simple form of this expression allows us to analytically calculate the local retarded Green's function in Eq. \eqref{eqn:Greensfunction_with_I}. For this driving, one has to distinguish between three time intervals when calculating the absolute value squared of $I\left(t_1,t_2\right)$.
If both the annihilation operator at $t_1$ and the creation operator at $t_2$ are applied before the field is switched on, i.e. $t_2<t_1<0$, the Hamiltonian equals a tight-binding Hamiltonian without an electric field as given in Eq. \eqref{eqn:tight_binding_H}. In this case, $I \left(t_1,t_2\right)=\trel$ and the local Green's function in Eq. \eqref{eqn:Greensfunction_with_I} is a Gaussian in $\trel$ multiplied by a step function, which becomes the Gaussian DOS after Fourier transformation to frequency \cite{Turkowski2005}. 
However, if the creation operator is applied before the field is switched on, meaning $t_2<0$, while the annihilation operator is applied after the field is turned on $\left(t_1>0\right)$, the absolute value of $I\left(t_1,t_2\right)$ is given by 
\begin{eqnarray}
\label{eqn:Ism_t<0}
&&\left|I_{\mathrm{sm}}^{t_2<0}\left(t_1,t_2\right)\right|^2=
\\
&&
\left|-t_2+
\cos\left(\frac{E_0}{\gamma}\right)\nonumber
F_c\left(t_1,0\right)
+
\sin\left(\frac{E_0}{\gamma}\right)
F_s\left(t_1,0\right)
\right|^2
\\
&&
+\left|
\cos\left(\frac{E_0}{\gamma}\right)
F_s\left(t_1,0\right)
-
\sin\left(\frac{E_0}{\gamma}\right)\nonumber
F_c\left(t_1,0\right)
\right|^2
.
\end{eqnarray}
Here $F_c$ and $F_s$ are the integrated functions $f_c\left(t\right)$ and $f_s\left(t\right)$ as defined in Eq. \eqref{eqn:integrands_f} and given by 
\begin{subequations}
\begin{eqnarray}
F_c\left(t_1,t_2\right)&=&
J_0\left(\frac{E_0}{\gamma}\right)\trel+\phi_c\left(t_1\right)-\phi_c\left(t_2\right)
\\
F_s\left(t_1,t_2\right)&=&
\phi_s\left(t_1\right)-\phi_s\left(t_2\right)
\,.
\end{eqnarray}
\end{subequations}
Finally, if the operator times obey $t_1>t_2>0$, the absolute square of $I\left(t_1,t_2\right)$ yields 
\begin{eqnarray}
\label{eqn:Ism_t>0}
&&\left|I_{\mathrm{sm}}^{t_2>0}\left(t_1,t_2\right)\right|^2=
\\
&&
\left|
\cos\left(\frac{E_0}{\gamma}\right)\nonumber
F_c\left(t_1,t_2\right)
+
\sin\left(\frac{E_0}{\gamma}\right)
F_s\left(t_1,t_2\right)
\right|^2
\\
&&
+\left|
\cos\left(\frac{E_0}{\gamma}\right)
F_s\left(t_1,t_2\right)
-
\sin\left(\frac{E_0}{\gamma}\right)\nonumber
F_c\left(t_1,t_2\right)
\right|^2
.
\end{eqnarray}
For large average times the DOS of the system coupling to the semi-infinite sinusoidal driving should equal the DOS of an infinite sinusoidal driving, and by factoring the expression in Eq. \eqref{eqn:Ism_t>0} it can indeed be shown that it is equal to the expression in 
Eq. \eqref{eqn:I_infinite_sin_driving_withbessel}, i.e. $\left|I_{\mathrm{sm}}^{t_2>0}\left(t_1,t_2\right)\right|^2=\left|I_\infty\left(t_1,t_2\right)\right|^2$. 
The function $F_c\left(t_1,t_2\right)$ is directly proportional to the Bessel function multiplied by $\trel$.
This means, that at large average times the relative time at which $t_2=\tave-\left(\trel/2\right)<0$ implies $\left|I_{\mathrm{sm}}^{t_2>0}\left(t_1,t_2\right)\right|^2$ in Eq. \eqref{eqn:Ism_t<0} is so large that the Green's function in Eq. \eqref{eqn:Greensfunction_with_I} is essentially zero. In this case, it does not contribute to the DOS anymore.

This holds true as long as $F_c\left(t_1,t_2\right)$ is indeed increasing with $\trel$, which is the case as long as the amplitude and the frequency of the driving are chosen in such a way that the Bessel function is not zero. However, if the Bessel function is zero, $\left|I_{\mathrm{sm}}^{t_2>0}\left(t_1,t_2\right)\right|^2$ is not increasing for increasing relative times, while $\left|I_{\mathrm{sm}}^{t_2<0}\left(t_1,t_2\right)\right|^2$ is still increasing because of the contribution of $-t_2$. This means that even for large average times, the Green's function at $t_2<0$, i.e. $\trel>2\tave$, contributes to the diagonal DOS, which will never be a set of delta functions and therefore never equal the DOS of the system coupled to an infinite drive. 

\begin{figure}
    \includegraphics[width= \columnwidth]{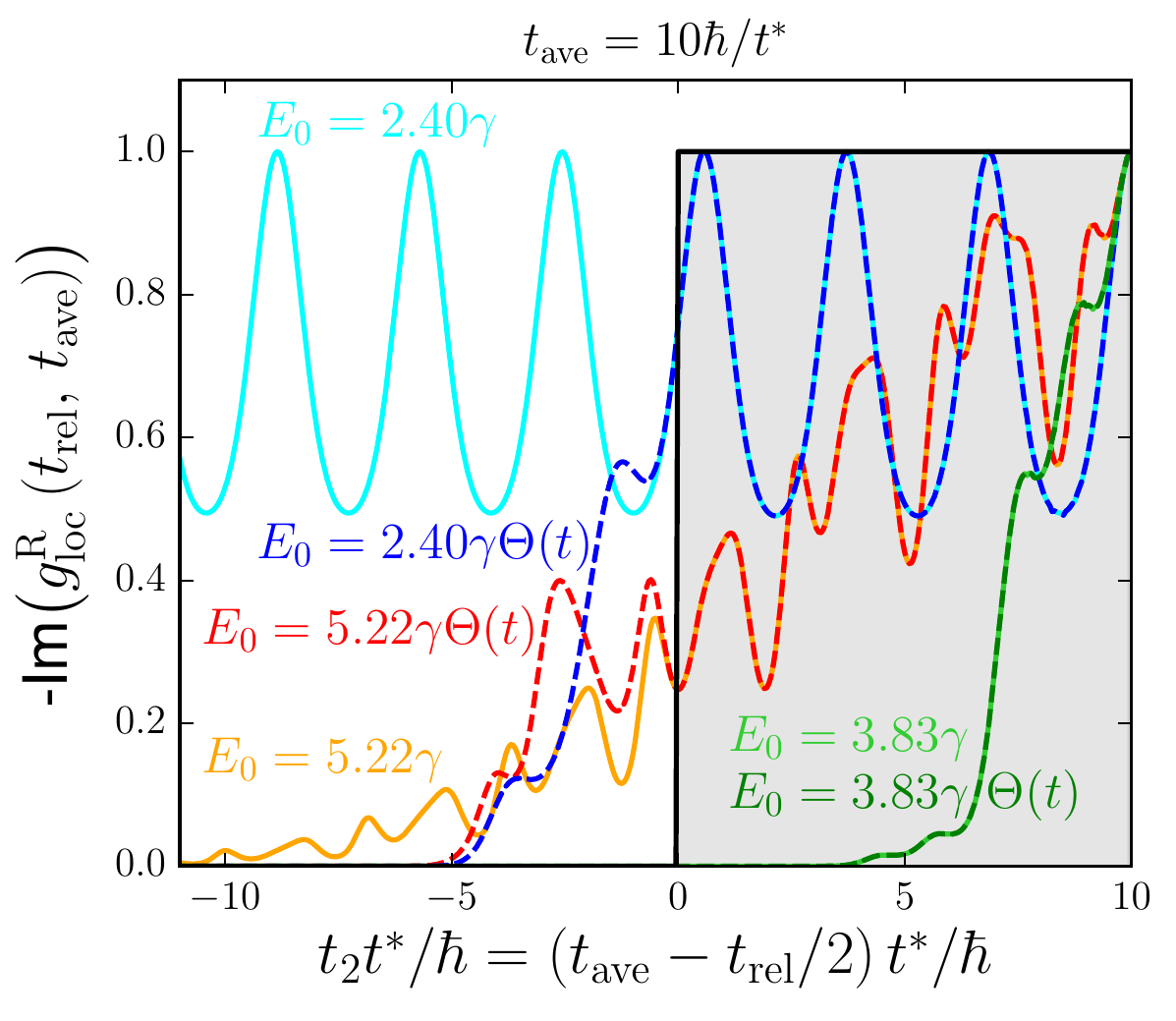}   
    \caption{
    (color online)
    Negative imaginary part of the local Green's function at $\tave=10\hbar/t^*$, as a function of $t_2$ in units of inverse rescaled hopping $\hbar /t^*$ for different electric fields. Other parameters: $\gamma=t^*/\hbar$. Note that the real part vanishes because $\mu=0$.}
    \label{fig:g_arg_t2}
\end{figure}
This scenario is displayed in Fig. \ref{fig:g_arg_t2}. Here the imaginary part of the local, time-dependent Green's function is plotted as a function of $t_2$ for fixed $\tave=10\hbar/t^*$. Note that for retarded quantities $t_2\leq\tave$ holds. The black line is the Heaviside step function, so the semi-infinite drive is turned on only in the area shaded in grey. The dashed lines correspond to the imaginary parts of the local, time-dependent Green's functions of a semi-infinite drive, while the solid lines correspond to the infinite drive. Within the shaded box, the curves at the same amplitude $E_0$ match perfectly. But outside of that area, where $t_2<0$, the imaginary parts of the local, time-dependent Green's functions corresponding to the semi-infinite drive decay 
faster than the functions corresponding to an infinite drive. 

For the amplitude $E_0=3.83\gamma$, both the Green's function for infinite drive (light green) and for semi-infinite drive (dark green) have completely decayed when $t_2=0$. This is because the magnitude of the Bessel function is large at $J_0\left(3.83\right)=-0.40$. In this case, choosing $\tave=10\hbar/t^*$ is sufficient to interpret the DOS corresponding to the     
semi-infinite drive as a Floquet DOS. Contrary to that, the imaginary part of the local, time-dependent Green's function with $E_0=5.22\gamma$ differs significantly from zero at $t_2=0$. This is because the magnitude of $J_0\left(5.22\right)=-0.10$ is small. For $t_2<0$ the function for semi-infinite drive (red) decays faster than for infinite drive (orange), so the DOS will not match due to these contributions from before $t=0$ (when the semi-infinite drive is turned off). At this amplitude, only the DOS corresponding to larger average times can approximate the Floquet results. Finally, the local time dependent Green's function corresponding to $E_0=2.40\gamma$ does not decay at all for the infinite drive (light blue). This is because $J_0\left(2.40\right)=0$. However, for the semi-infinite drive (dark blue) it starts to decay immediately for $t_2<0$. This means the DOS corresponding to the infinite drive and the semi-infinite drive will never match, no matter how large the average time is chosen to be, and the DOS corresponding to semi-infinite driving can never be interpreted as the Floquet DOS.

However, for the horizontal Fourier transformation the Green's function only contributes to the horizontal DOS for $t>t_2$, which is why the horizontal DOS corresponding to the semi-infinite sinusoidal drive always equals the horizontal DOS corresponding to the infinite drive for all $t_2>0$ for any amplitude and frequency of the electric field.
\subsection{Sinusoidal Steplike Pulse}
A pulse that is turned on at $t_0$ and turned off after $n\in\mathbb{N}$ oscillations, i.e. at a cutoff time $t_\mathrm{c}=2\pi n/\gamma$ is not experimentally implementable either, but there are experimental implementations that come close. One advantage of it is, that again the DOS can be computed analytically. 
Naturally, for $t_2<t_1<\tc$ the DOS equals the results for semi-infinite driving, i.e. depending on the sign of $t_2$ the absolute value of $I\left(t_1,t_2\right)$ is given by Eq. \eqref{eqn:Ism_t<0} or Eq. \eqref{eqn:Ism_t>0}. 
This also means that for the diagonal Fourier transform, the average time needs to be chosen large enough for the DOS corresponding to the semi-infinite driving to equal the DOS  obtained by applying an infinite sinusoidal driving, as explained in section \ref{subsec:semi_infinite_driving}. 
However, for $0<t_2<\tc<t_1$, the absolute value squared of $I\left(t_1,t_2\right)$ is given by
\begin{eqnarray}
\label{eqn:Ism_t2<tE}
&&\left|I_{\mathrm{sp}}^{t_c<t_1}\left(t_1,t_2\right)\right|^2=
\\
&&
\left|t_1-\tc+
\cos\left(\frac{E_0}{\gamma}\right)\nonumber
F_c\left(\tc,t_2\right)
+
\sin\left(\frac{E_0}{\gamma}\right)
F_s\left(\tc,t_2\right)
\right|^2
\\
&&
+\left|
\cos\left(\frac{E_0}{\gamma}\right)
F_s\left(\tc,t_2\right)
-
\sin\left(\frac{E_0}{\gamma}\right)\nonumber
F_c\left(\tc,t_2\right)
\right|^2
,
\end{eqnarray}
where the growing $t_1-\tc$ for increasing $\trel$ causes significant deviations from the DOS corresponding to infinite driving. Therefore, it is not enough to choose $\tave$ to be large and the Bessel function to have a finite size in order to interpret the results with Floquet theory, but also $\tc$ must be large enough that the change in the local Green's function that is caused by Eq. \eqref{eqn:Ism_t2<tE} has no further effect on the DOS. 
Similar to the semi-infinite driving, the dephasing in $\trel$ of the local retarded Green's function is significantly faster if the Bessel function of the amplitude divided by the frequency of the electric field is large. 
Since the absolute value of $I\left(t_1,t_2\right)$ for $t_2<0$ and for $t_1>\tc$, given in  Eq. \eqref{eqn:Ism_t<0} and Eq. \eqref{eqn:Ism_t2<tE} respectively, is eventually increasing for any electric field, 
while it is oscillating equally to the absolute value of $I\left(t_1,t_2\right)$ of the infinite sinusoidal driving  for $0<t_2<t_1<\tc$ and $J_0\left(E_0/\gamma\right)=0$, the DOS will never be a set of delta functions. The Green's function for $t_1>\tc$ will eventually contribute to the DOS for both the horizontal and the diagonal Fourier transform, therefore at zeros of the Bessel function, the measured DOS corresponding to the sinusoidal steplike pulse can never be interpreted using Floquet theory, no matter which Fourier transform is chosen. 

Studying the diagonal DOS for $0<\tave<\tc/2$, it is obvious that $t_2<0$ applies before $t_1>\tc$ needs to be taken into account, therefore here the results correspond to the results for the semi-infinite driving. 
However, for $\tc/2<\tave<\tc$, the Green's function at $\tc<t_1$ needs to be considered before $t_2<0$ applies. Therefore, for Floquet theory to be valid, the Green's function has to be approximately zero at $t_1=t_c$ in order for it to have a negligible contribution to the DOS. 
Because the dephasing in $\trel$ in the local Green's function is faster the larger the value of the Bessel function multiplied by $\sin\left(E_0/\gamma\right)$ and $\cos\left(E_0/\gamma\right)$ is, the cutoff time $t_c$ can be chosen significantly smaller for large values of the Bessel functions. 
\begin{figure}
    \includegraphics[width= \columnwidth]{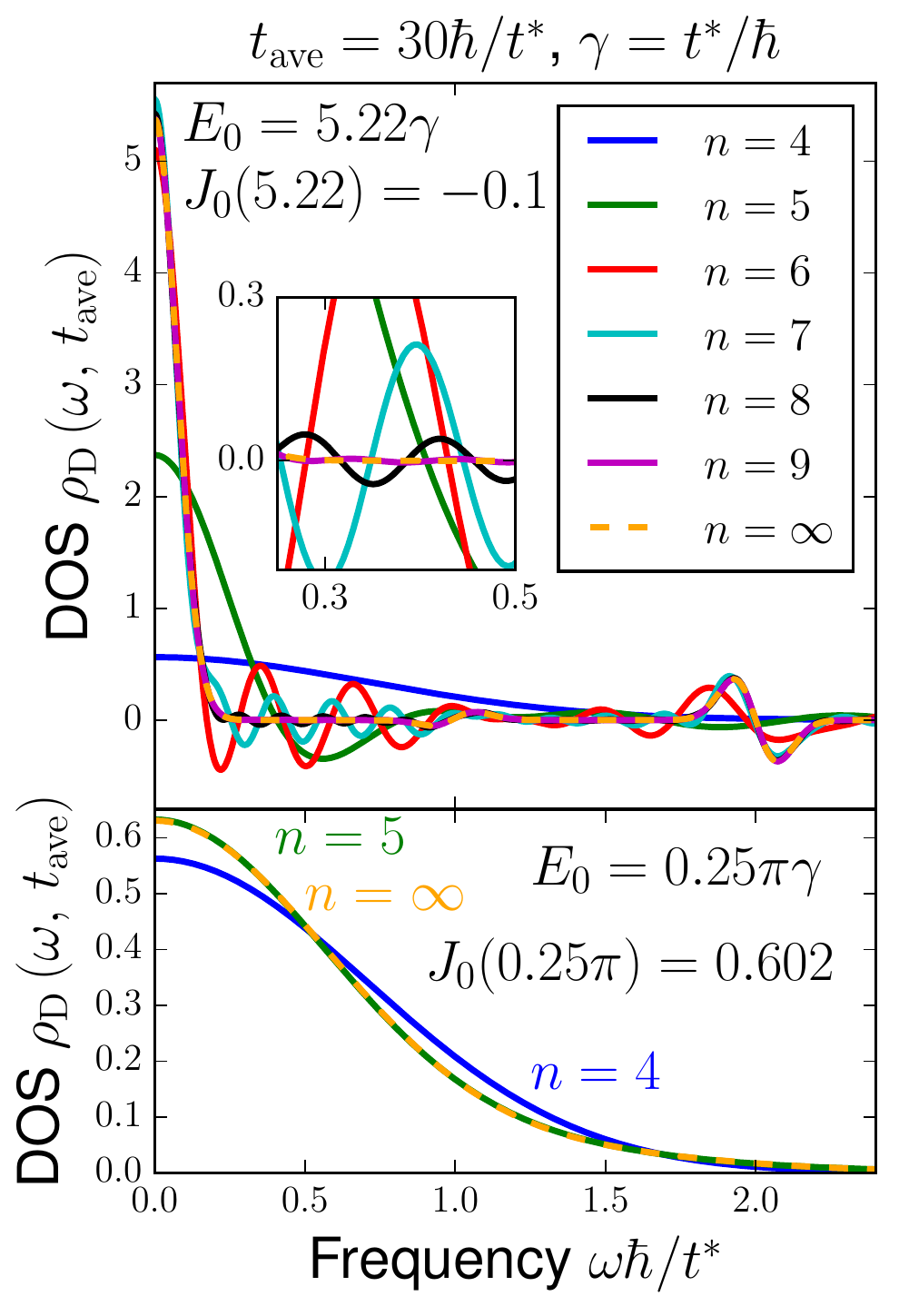}    
    \caption{(color online) The diagonal DOS corresponding to the steplike pulse that starts at $t=0$ and is turned off after $n$ oscillations at a cutoff time $\tc=2\pi n/\gamma$ and corresponding to an infinite sinusoidal drive for $\gamma=t^*/\hbar$, $\tave=30\hbar/t^*$ and two different amplitudes $E_0=5.22\gamma$ (upper panel) and $E_0=\pi\gamma/4$ (lower panel).}
    \label{fig:sin_steplike_pulse}
\end{figure}

Note that it is most suitable to set the average time to $\tave=\tc/2$ because in this case $t_2=0$ and $t_1=t_c$ occur at the same relative time $\trel=\tc$. When $\tave$ is chosen to be the minimal average time $\tave^{\mathrm{min}}$ at which the DOS corresponding to a semi-infinite drive equals the DOS for infinite driving, then $\tc=2\tave^{\mathrm{min}}$ is the shortest cutoff time at which the DOS of the steplike pulse can be interpreted with Floquet theory.

Figure \ref{fig:sin_steplike_pulse} displays the  diagonal DOS corresponding to the steplike pulse and to infinite sinusoidal driving for $\gamma=t^*/\hbar$ at two different amplitudes of the electric field. The average time is chosen to be $\tave=30\hbar/t^*$ because earlier analyses have shown that this is sufficiently large for the diagonal DOS of the semi-infinite sinusoidal drive to equal the DOS associated with infinite driving. 
For $t_c<\tave$, the squared absolute value of $I\left(t_1,t_2\right)$ is given by $\left|\trel\right|^2$, so for $t_2<t_1<0$, the local retarded Green's function (and therefore the DOS) are Gaussian and equal to the noninteracting DOS, as explained in section \ref{subsec:semi_infinite_driving}. This is why the blue line at $n=4$, i.e. at $\tc=4\cdot 2\pi\hbar/t^*<\tave=30\hbar/t^*$, is Gaussian and the same for both amplitudes of the electric field. 

In the upper panel of Fig.~\ref{fig:sin_steplike_pulse}, the amplitude of the electric field is given by $E_0=5.22\gamma$, so that the magnitude of the Bessel function is small $\left[J_0\left(5.22\right)=-0.10\right]$. This means the dephasing in $\trel$ is slow (as can be seen in Fig. \ref{fig:infinite_sinusoidal}), so $t_c$ has to be chosen quite large in order for Floquet theory to be applicable. When contemplating the DOS at $n\in\left\lbrace 5,6,7\right\rbrace$ in the upper panel, it is obvious that it is completely different from the DOS of the infinite sinusoidal pulse, which is displayed in orange. 
Only at $n=8$ do we start to see some similarity, and the lines seem to match at $n=9$. However, only $n=10$ (this is not displayed, as the deviations from $n=9$ are too small to be seen) is sufficient for the two diagonal DOS to be essentially equal. This means $t_c$ needs to be chosen to be twice as large as $\tave$ when the results are meant to be interpreted with Floquet theory. 

In the lower panel of Fig.~\ref{fig:sin_steplike_pulse}, the amplitude of the electric field is chosen to be $E_0=0.25\pi\gamma$. In this case, the magnitude of $J_0\left(0.25\pi\right)=0.602$ is large and both $\cos\left(0.25\pi\right)=\sin\left(0.25\pi\right)=1/\sqrt{2}$ are large too. Contrary to the DOS corresponding to a small value of the Bessel function, we find that the Gaussian diagonal DOS  at $\tave>\tc$ ($n=4$, blue), shows similarities to the diagonal DOS corresponding to the infinite sinusoidal driving. Furthermore, as soon as $\tave<\tc$ ($n=5$, green), the DOS are equal. Note that the average time in the lower panel is chosen to be $\tave=30\hbar/t^*$ to ensure comparability with the upper panel. 
But while $\tave=30\hbar/t^*\approx \tave^\mathrm{min}$ holds for $E_0=5.22\gamma$ (upper panel), the minimal average time for $E_0=0.25\pi\gamma$ is much smaller at $\tave^\mathrm{min}\ll\tave=30\hbar/t^*$. Therefore, in the lower panel, $\tc\ll2\tave$ is sufficient to interpret the DOS with Floquet theory.

The observations above hold for the horizontal DOS as well, the major difference being that $\tave$ does not need to be chosen as large. In fact, in this case, choosing $\tave=0$ is ideal, as only the magnitude of $\tc-\tave$ determines the quality of the results for a given electric field.
\subsection{Sinusoidal Gaussian Pulse}
\begin{figure}
    \includegraphics[width=0.95\columnwidth]{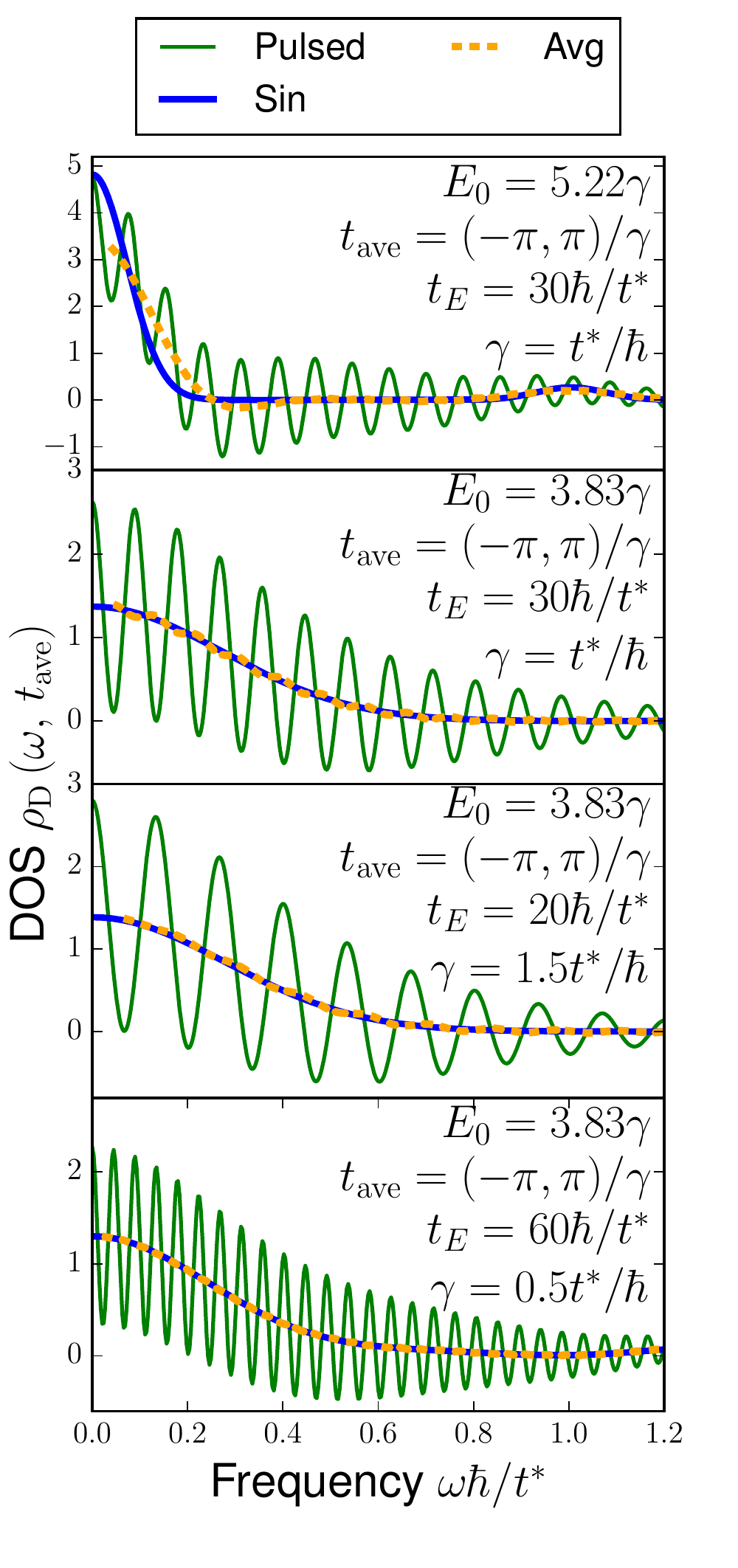}
      \caption{(color online) The diagonal DOS corresponding to an infinite sinusoidal drive (blue) and a pulsed system (green), both averaged over the Floquet period from $\tave=-\pi/\gamma$ to $\tave=\pi/\gamma$ and the running average in frequency $\omega$ (orange) over one period of the oscillations in the diagonal DOS of the pulsed system at different amplitudes and frequencies of the electric field and for a Gaussian pulse of varying width.}    
    \label{fig:DOS_Gaus_newFourier_even}
\end{figure}

A field pump that is implementable in an experiment is a sinusoidal electric field that is modulated with a Gaussian envelope, i.e. an electric field that is given by 
\begin{equation}
\label{eqn:gaussian_e_field}
E\left(t\right)=E\sin\left(\gamma t\right)
\mathrm{exp}\left[-\left(\frac{t}{\te}\right)^2\right]\,,
\end{equation}
where $\te$ is the width of the Gaussian. This corresponds to the vector potential (expressed in terms of the imaginary error function in the first line and the Faddeeva function in the second line)
\begin{subequations}
\label{eqn:gaussian_vector_potential}
\begin{eqnarray}
A_\mathrm{G}\left(t\right)=
\frac{cE\te \sqrt{\pi}}{2}
e^{-\left(\frac{\gamma \te}{2}\right)^2}
\mathrm{Re}\left[
\mathrm{erfi}\left(
\frac{\gamma\te}{2}+\frac{it}{\te}\right)
\right]\\
=
\frac{cE\te \sqrt{\pi}}{2} e^{-\left(\frac{t}{\te}\right)^2}
\mathrm{Im}\left[
\mathrm{w}\left(
\frac{\gamma\te}{2}+\frac{it}{\te}\right)e^{i\gamma t}
\right],\,
\end{eqnarray}
\end{subequations}
which shares many properties with the vector potential for the infinite sinusoidal pulse in Eq. \eqref{eqn:inf_vector_potential}. Both vector potentials have the same zeros and oscillate with the same frequency, the major difference is the decaying amplitude for the oscillations in the vector potential $A_\mathrm{G}$. 
Since we are not able to calculate an analytic form for the local retarded Green's function from that, we simply calculate it numerically. What we find is that as the Gaussian broadens, the corresponding DOS does not simply approach the DOS of an infinite sinusoidal drive, but oscillates around it (see Fig. \ref{fig:DOS_Gaus_newFourier_even}). 
It requires a second averaging, more precisely a running average over one period of these oscillations in $\omega$, for the DOS of the Gaussian pulse to match the DOS of the infinite sinusoidal driving. 

\begin{figure}
    \includegraphics[width= \columnwidth]{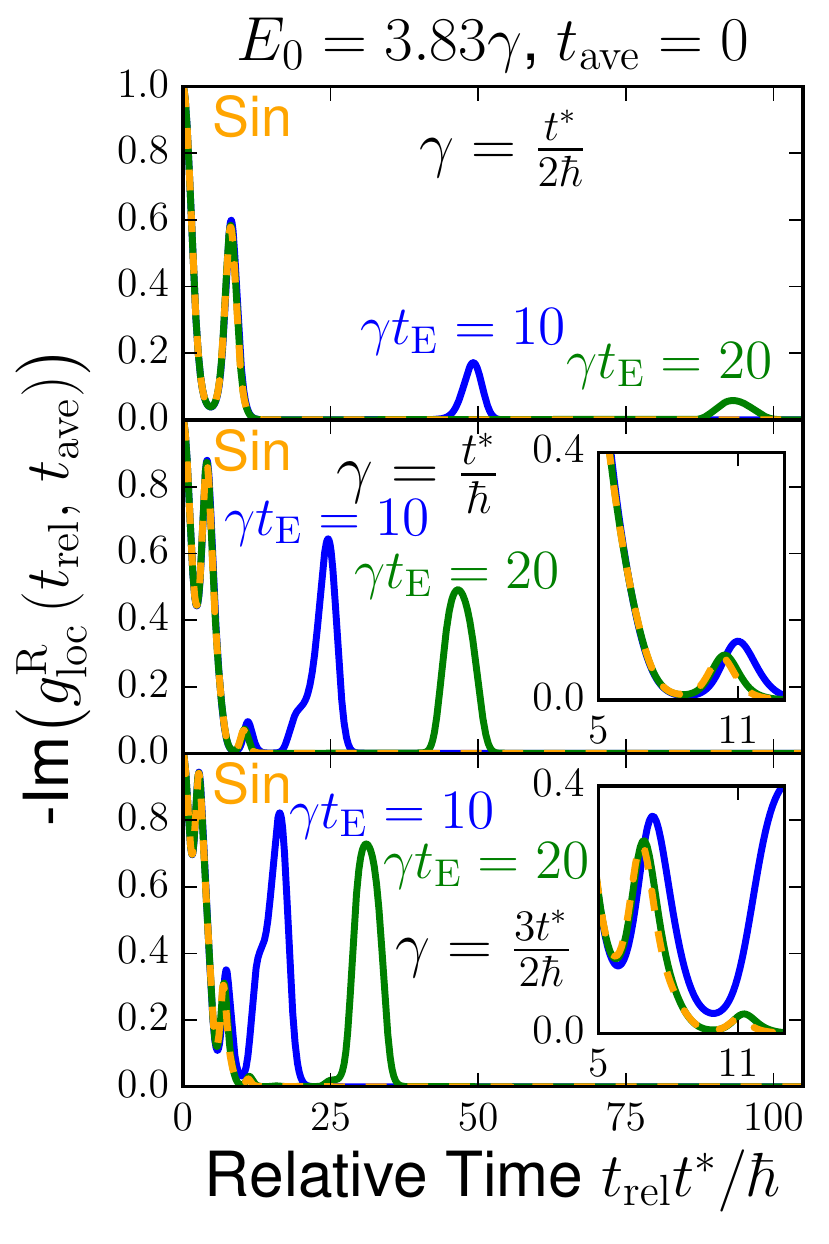}    
    \caption{(color online)  Negative imaginary part of the local Green's function at a constant average time $\tave=0$, as a function of $\trel$ in units of inverse rescaled hopping $\hbar/t^*$ at $E_0=3.83\gamma$ for an infinite sinusoidal drive (orange) and for two Gaussians where the product of the width of the Gaussian and the frequency of the electric field is $\gamma\te=10$ (blue) and $\gamma\te=20$ (green) respectively, at three different driving frequencies $\gamma=0.5t^*/\hbar$ (upper panel),  $\gamma=t^*/\hbar$ (middle panel) and  $\gamma=1.5t^*/\hbar$ (lower panel).}
    \label{fig:g_Arg_Emin_varygamma}
\end{figure}

This can be traced back to an additional peak in the imaginary part of the time-dependent local Green's function, whose position and shape are functions of the amplitude $E_0$, the frequency $\gamma$ and the width $\te$ of a pulse (see Fig. \ref{fig:g_Arg_Emin_varygamma}). 
Figure \ref{fig:g_Arg_Emin_varygamma} shows that for a wide Gaussian, the imaginary part of the time-dependent local Green's function perfectly matches the imaginary part of the time-dependent local Green's function of the system coupling to an infinite drive up to a relative time at which the Green's function of the infinite drive completely decays. 
The Green's function of the Gaussian pulse has a single complex revival at a later relative time.
 
The amplitude of the electric field is chosen to be $E_0=3.83\gamma$ so the Bessel function of the amplitude divided by the frequency is at its first minimum. This ensures a fast decay of the Green's function corresponding to the infinite sinusoidal drive as explained in Sect.~\ref{subsec:infinite_driving}. Therefore, this amplitude leads to the best agreement between the Green's function corresponding to an infinite drive (orange) and the Green's function of the pulsed systems (blue and green) before the latter Green's functions have their revival. 

This behavior is illustrated further in Fig. \ref{fig:DOS_Gaus_newFourier_even}, which displays the diagonal DOS corresponding to an infinite sinusoidal drive (blue) and a pulsed system (green) in frequency space. Both diagonal DOS are averaged over the Floquet period from $\tave=-\pi/\gamma$ to $\tave=\pi/\gamma$, evenly around the center of the pulse. 
Any averaging that is not centered around the pulse leads to significantly worse results. 
The orange line in Fig. \ref{fig:DOS_Gaus_newFourier_even} is the running average in frequency $\omega$ over one period of the oscillations in the diagonal DOS of the pulsed system, i.e., the average over one period of the oscillations of the green line. 

For an amplitude at which the Bessel Function $J_0\left(E_0/\gamma\right)$ is small ($E_0=5.22\gamma$, upper panel in Fig. \ref{fig:DOS_Gaus_newFourier_even}), the diagonal DOS of the pulsed system shows large deviations from the diagonal DOS corresponding to the infinite sinusoidal drive even after taking the running average (especially for small frequencies $\omega$). 
However, at $E_0=3.83\gamma$ (first minimum of the Bessel function, second panel in Fig. \ref{fig:DOS_Gaus_newFourier_even}), the agreement between the diagonal DOS of the system coupling to an infinite sinusoidal drive and the running average over the diagonal DOS of the pulsed system is good (all parameters except for $E_0$ are the same in the two upper panels). 

To explain the connection between the frequency of the oscillations in the DOS and both the width of the Gaussian $\te$ and the frequency $\gamma$ of the electric field, it is useful to study the imaginary part of the local time dependent Green's function (see Fig. \ref{fig:g_Arg_Emin_varygamma}). 
For a wider Gaussian, i.e., larger $\te$ (for fixed $\gamma$) the revival occurs later, meaning that the oscillations in the DOS show a higher frequency. In fact, the time at which the revival occurs seems to be almost linearly connected to the width of the Gaussian, as a shift by some factor $\alpha$ in the width $\te\rightarrow \alpha\te$ leads to the revival time shifting from $\trel$ to $\alpha\trel$. 
Varying $\gamma$ on the other hand has little effect on the relative time at which the revival occurs, but for a constant pulse width $\te$ the agreement between the Green's function of the infinitely driven system and the Green's function of the pulsed system diminishes for very small $\gamma$. 
Another disadvantage of low frequencies is that the Fourier period $2\pi/\gamma$ increases, so when calculating the averaged DOS the Green's function requires contributions from average times that are much further away from the center of the pulse.

Figure \ref{fig:g_Arg_Emin_varygamma} shows that at any given frequency $\gamma$, the revival occurs later whenever the product $\gamma\te$ is larger.
That is, the green peak for $\gamma\te=20$ always occurs at a later relative time than the blue peak at $\gamma\te=10$.
The agreement between the running average of the DOS of the pumped system and the DOS of the system coupling to an infinite drive is generally better for later times of the revival in the Green's function. 
This is because for early arrival times, the Green's function corresponding to the infinite drive may not have completely decayed when the revival occurs. That is, the resemblance between the two DOS is better at high frequencies $\gamma$ and for broad Gaussian pulses.  
This is not surprising, as it means that the pumped field resembles the infinite sinusoidal field when it has a larger amount of oscillations. 
Therefore it is more interesting to compare the Green's functions and the resulting DOS at varying frequencies $\gamma$ where the product $\gamma\te$ of the width of the pulse and the frequency of the electric field is kept constant.

By comparing the revival times at the frequencies $\gamma=0.5t^*/\hbar$, $\gamma=t^*/\hbar$ and $\gamma=1.5t^*/\hbar$ (with $\gamma\te$ fixed) in Fig. \ref{fig:g_Arg_Emin_varygamma}, it is clear that the revival occurs later for lower frequencies. 
This directly results from the later occurrence of the revival as the Gaussian broadens.  
At $\gamma=0.5t^*/\hbar$ (upper panel of Fig. \ref{fig:g_Arg_Emin_varygamma}), the revivals of both $\gamma\te=10$ and $\gamma\te=20$ occur long after the Green's function corresponding to the infinite drive has decayed. This means that the agreement between the curves is good up to this point. But this agreement becomes worse at larger frequencies. For $\gamma=t^*/\hbar$ (middle panel in Fig. \ref{fig:g_Arg_Emin_varygamma}) the green curve at $\gamma\te=20$ still matches the Green's function of the system coupling to an infinite drive up to the point where the latter one has decayed, but the revival of the blue curve at $\gamma\te=10$ moves to times $\trel$ where the Green's function corresponding to the infinite drive has not completely decayed. The deviations before the decay of the Green's function become even larger at higher frequencies like $\gamma=1.5t^*/\hbar$ (lower panel in Fig. \ref{fig:g_Arg_Emin_varygamma}).
This implies that the applicability of Floquet theory is strongly dependent on the width of the Gaussian, and to a lesser extent on the driving frequency $\gamma$. A wide Gaussian ensures that the measured DOS resembles the DOS of the infinitely driven system, even if the frequency of the driving field is low. 

Figure \ref{fig:DOS_Gaus_newFourier_even} confirms these conclusions. Comparing the lower three panels, where the width of the Gaussian and the frequency of the electric field are chosen so $\gamma\te=30$ holds, it becomes clear that the frequency with which the diagonal DOS of the pulsed system oscillates around the DOS corresponding to the infinite drive is increasing as the Gaussian broadens and the frequency of the driving field decreases. Note that even though the diagonal DOS of the pulsed system (green) is averaged over the Floquet period, the oscillations take negative values, i.e. the averaged DOS is not semidefinite even if the Gaussian is broad. 
As explained in Sect.~\ref{subsec:infinite_driving}, it is required for the DOS averaged over the Floquet period to be semidefinite if we want the pulsed system to be representative of the Floquet results. 
Fortunately, the orange line that results from calculating the running average over one period of the oscillations in the DOS (corresponding to the Gaussian pulse) is semidefinite and resembles the DOS of the infinitely driven system well.
\begin{figure}
    \includegraphics[width= \columnwidth]{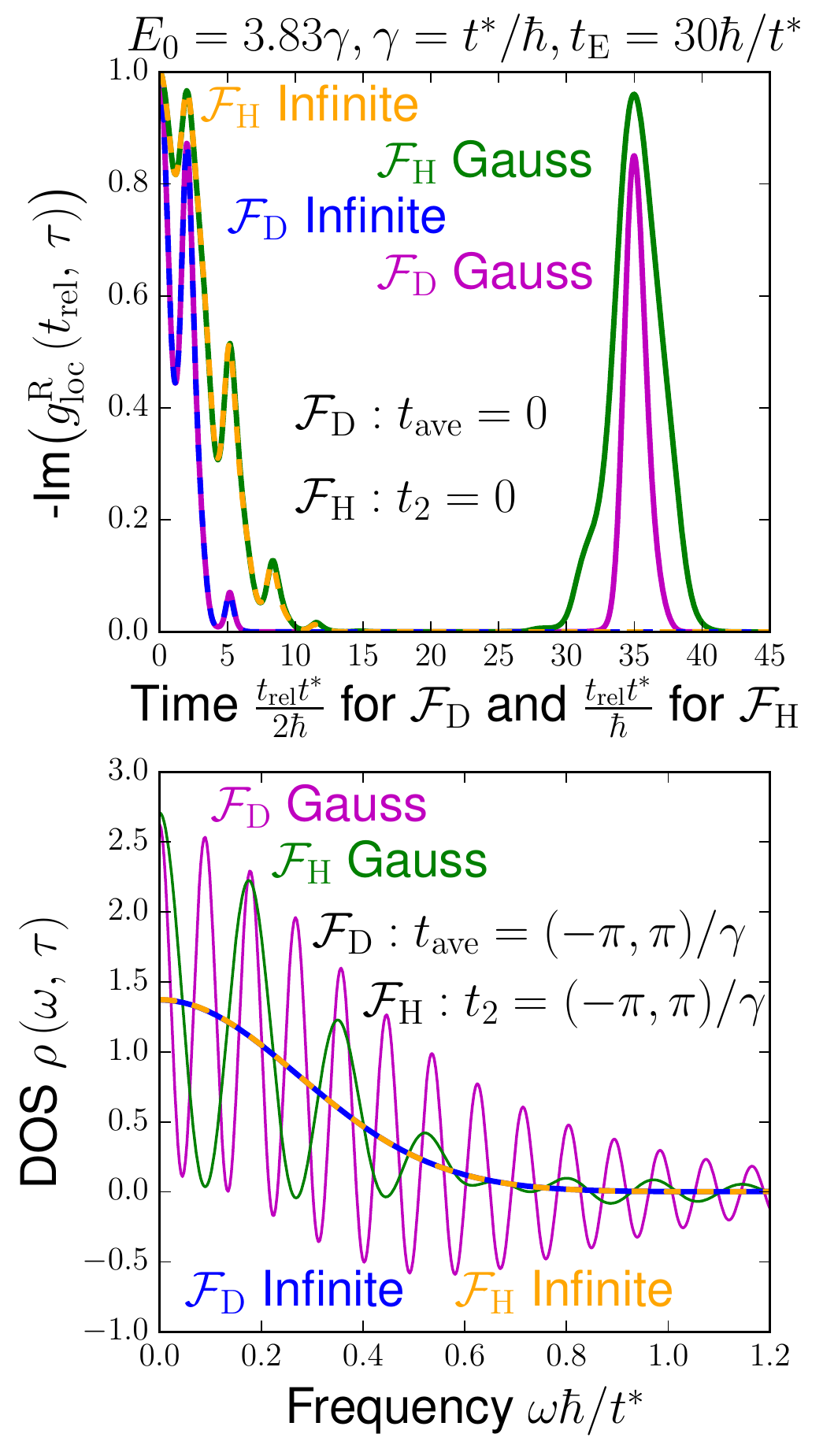}    
    \caption{(color online) The upper panel: Green's function of both the pulsed system (labeled "Gauss") and the system that is driven for an infinitely long time (labeled "Infinite") for a constant $t_2=0$ ($\mathcal{F}_\mathrm{H}$) and a constant $\tave=0$ ($\mathcal{F}_\mathrm{D}$). Lower panel: horizontal and diagonal DOS of both systems, averaged over one Fourier period in $t_2$ and $\tave$ respectively. Other parameters: $E_0=3.83\gamma$, $\gamma=t^*/\hbar$ and $\te=30\hbar/t^*$.}
    \label{fig:diff_fourier_transformations}
\end{figure}

The results for the horizontal Fourier transformation are similar to the results obtained using the diagonal Fourier transformation. The major difference is a factor of $2$ in the relative time that was already mentioned in Sec. \ref{subsec:infinite_driving} and is caused by the relation $t_1=\tave+\trel/2$. 
The upper panel in Fig. \ref{fig:diff_fourier_transformations} shows the Green's function of both the pulsed system and the system that is driven for an infinitely long time at $E_0=3.83\gamma$, $\gamma=t^*/\hbar$ and $\te=30\hbar/t^*$, both for $\tave=0$ and for $t_2=0$. Note that the Green's functions are given in terms of $\trel/2$ for the constant average time $\tave=0$ (magenta) and in terms of $\trel$ for a constant $t_2=0$ (green). 
This is to emphasize the fact that the peak of the revival in the Green's function corresponding to a constant $t_2$ takes place at exactly half of the relative time at which the peak of the revival in the Green's function corresponding to a constant $\tave$ is located. 
In fact, for small relative times, even the relative times at which the Green's functions of the system that is coupling to an infinite drive (orange and blue) go through extrema are separated by this factor $2$. 

The lower panel of Fig.~\ref{fig:diff_fourier_transformations} displays the diagonal and the horizontal DOS of the pulsed system and the infinitely driven system, both averaged over one Fourier period from $\tave=-\pi/\gamma$ to $\tave=\pi/\gamma$ for constant $\tave$ and from $t_2=-\pi/\gamma$ to $t_2=\pi/\gamma$ for constant $t_2$. 
Note that this averaging leads to exactly the same DOS for the infinite drive (blue and orange), no matter which Fourier transform is computed (as shown in Sec. \ref{subsec:infinite_driving}).  The diagonal DOS (magenta) of the pulsed system, however, oscillates with almost double the frequency of the horizontal DOS (green) of the same system. 
While the period of the oscillations in the horizontal DOS is almost perfectly constant and the amplitude of these oscillations decreases monotonically, the period of the oscillations of the horizontal DOS varies significantly more. Though the amplitude of the horizontal DOS shows an overall decay, it does not decrease monotonically. 
This leads to a slightly worse agreement between the DOS corresponding to the infinite drive and the running average over one period of the oscillations in the DOS for the Gaussian pulse when the DOS is horizontal.

Note that the second "revival" peak of the time dependent local Green's function is expected to be smaller for interacting systems where Green's functions decay more rapidly in imaginary time. So these oscillations may be reduced when interactions are included.

\section{Summary}

In this work, we have examined situations where one might be able to observe Floquet behavior.
We studied noninteracting, fermionic systems (which do not heat up) and compared the exact Floquet solutions for the retarded Green's functions to a number of different cases including a semi-infinite drive and a periodic drive with a Gaussian envelope (to make it into a pulse which is experimentally realizable). The true Floquet system has a Hamiltonian that is periodic with respect to the period of the driving. 

We observed a number of interesting results. First,  for the pure Floquet system, the conventional definition of the instantaneous DOS as the imaginary part of the retarded local  Green's function evaluated at fixed $\tave$ or $t_2$, is not positive semi-definite. But the time-averaged density of states always is. This holds both for the diagonal and the horizontal Green's function.

Second, when an ac electric field is applied along the main diagonal direction of the lattice, the value of the Bessel function $J_0\left(E_0/\gamma\right)$ is critical in determining the subsequent behavior. In the Floquet limit, one will obtain a local DOS that is a sequence of delta functions when $J_0\left(E_0/\gamma\right)=0$; they become broadened and lose their identity as $J_0$ becomes larger in magnitude.

Third, even if the Hamiltonian is not strictly periodic, Floquet theory is still applicable as a good approximation if certain other requirements are met. In particular, when  $\left|J_0\left(E_0/\gamma\right)\right|$ is large, the pulsed system appears quite close to the Floquet result. But, as mentioned above, when we are at a zero of the Bessel function, it is never feasible to find the pumped system looking like the Floquet one.

In particular, if we employ a Gaussian envelope function, the width of the envelope is the primary
predictor of whether the system will look like a periodic Floquet system.  A wide Gaussian ensures that the measured DOS resembles the DOS of the infinitely driven system, even if the frequency of the driving field is low. On the other hand, measuring at high frequencies does not compensate for a narrow Gaussian. Hence, it is not true that one can simply count the number of oscillations inside one or two standard deviations of the pulse to determine whether it will behave like a Floquet system---this only holds if the Gaussian pulse width is wide enough. 

Surprisingly, even if the system resembles a periodic Floquet system in the time domain, it is not sufficient to average the DOS over one period of the driving (in $\tave$ or $t_2$) to reproduce the DOS of the corresponding Floquet Hamiltonian (even if the amplitude $E_0$, the frequency $\gamma$ and the width of the Gaussian $t_E$ are optimally chosen). Instead, it further requires a second averaging, precisely the running average (in the frequency domain) over one period of the oscillations , for the pulsed DOS to resemble the DOS of the infinite drive.

As interactions are added in (see as a first step Ref.~\cite{Kennes2018}), we expect it to be easier for the Gaussian pumped system to look Floquet like, because the extra scattering due to the interactions will cause the Green's functions to decay more rapidly in relative time. This will, in turn, widen the parameter space where the pulsed system appears to behave more like the corresponding Floquet system.  
If the pump adds substantial heat to the system a high temperature stationary state will be reached in which we do not expect the retarded Green's function to depend strongly on temperature.
Of course it will have larger effect on lesser Green's functions, but we are not discussing those here.
We look forward to seeing more experiments that will illustrate this behavior in the future.

\begin{acknowledgments}
One of us (MHK) acknowledges the financial support by the Studienstiftung des Deutschen Volkes. This work was supported by the Department of Energy, Office of Basic Energy Sciences, Division of Materials Sciences and Engineering (DMSE) under contract No. DE-FG02-08ER46542 (JKF) and by the Deutsche Forschungsgemeinschaft in project
UH 90-13/1 (GSU). JKF also acknowledges support from the McDevitt Bequest at Georgetown. 
\end{acknowledgments}

\appendix
\section{Semipositivity of the Time-Averaged Densities of States}
\label{sec:Appendix}
We start with two arbitrary $2\pi/\gamma$ periodic functions $f\left(t\right)$ and $g\left(t\right)$, which  can be expressed as the following Fourier series:
\begin{subequations}
\begin{eqnarray}
f\left(t\right)&=&\sum_m e^{im\gamma t} f_m\\
g\left(t\right)&=&\sum_m e^{im\gamma t} g_m\,.
\end{eqnarray}
\end{subequations}
The convolution of these two functions is given by
\begin{subequations}
\begin{eqnarray}
h\left(t\right)&=&\frac{\gamma}{2\pi}\int_x^{x+\frac{2\pi}{\gamma}}
g\left(t-t'\right)f\left(t'\right)\mathrm{d}t'
\\
&=&
\frac{\gamma}{2\pi}\int_x^{x+\frac{2\pi}{\gamma}}
\sum_{m,n} 
e^{in\gamma \left(t-t'\right)} e^{im\gamma t'} g_n f_m\mathrm{d}t'
\\
&=&\sum_{m,n}
g_n f_m e^{int\gamma}
\underbrace{\frac{\gamma}{2\pi}\int_x^{x+\frac{2\pi}{\gamma}}
e^{i\gamma\left(m-n\right)t'}
\mathrm{d}t'}_{\delta_{m,n}}\\
&=&
\sum_{m}
g_m f_m e^{imt\gamma}
\\
&=&
\sum_{m}
h_m e^{imt\gamma}\,.
\end{eqnarray}
\end{subequations}
which is also $2\pi/\gamma$ periodic and has Fourier coefficients
$h_m=g_m f_m$. 
This means that if the coefficients $f_m$ and $g_m$ are complex conjugates of each other, the coefficients of the convolution are positive and obey $h_m=\left|f_m\right|^2\geq 0$.
Coefficients that are complex conjugates naturally arise when the $2\pi/\gamma$ periodic functions obey $g\left(t\right)=f^*\left(-t\right)=\sum_m \mathrm{exp}\left[im\gamma t\right]f^*_m$.
Using this identity and substituting either $\tilde{t}=t'-t$ or $\tilde{\tilde{t}}=t'-\left(t/2\right)$, the convolution becomes 
\begin{subequations}
\begin{eqnarray}
h\left(t\right)&=&
\frac{\gamma}{2\pi}\int_x^{x+\frac{2\pi}{\gamma}}
f^*\left(t'-t\right)f\left(t'\right)\mathrm{d}t'
\\
&=&
\frac{\gamma}{2\pi}\int_{\tilde{x}}^{\tilde{x}+\frac{2\pi}{\gamma}}
f^*\left(\tilde{t}\right)f\left(\tilde{t}+t\right)\mathrm{d}\tilde{t}\label{eqn:conv_vert}
\\
&=&
\frac{\gamma}{2\pi}\int_{\tilde{\tilde{x}}}^{\tilde{\tilde{x}}+\frac{2\pi}{\gamma}}
f^*\left(\tilde{\tilde{t}}-\frac{t}{2}\right)
f\left(\tilde{\tilde{t}}+\frac{t}{2}\right)\mathrm{d}\tilde{\tilde{t}}\label{eqn:conv_diag}
\\
&=&
\sum_{m}
|f_m|^2 e^{imt\gamma}\,.
\end{eqnarray}
\end{subequations}
The averaged local retarded Green's function, as defined in Eq. \eqref{eqn:averaged_Greensfunction}
 of the infinitely driven field, has exactly the form of Eq. \eqref{eqn:conv_vert} when the retarded Green's function is given as a function of $t_1$ and $t_2$ by identifying $t_2=\tilde{t}$. At the same time it has exactly the form of \eqref{eqn:conv_diag} when writing the retarded Green's function as a function of $\tave$ and $\trel$ and identifying $\tave=\tilde{\tilde{t}}$. Therefore the time-averaged local retarded Green's function is given by 
 \begin{subequations}
 \begin{eqnarray}
 \bar{g}^R\left(\vect{k},\trel\right)&=&
 -\frac{i}{\hbar}\Theta\left(\trel\right)
 e^{-\frac{i\varepsilon\left(\vect{k}\right)}{\hbar}J_0\left(\frac{E_0}{\hbar}\right)\trel}
\\
&&\times \sum_m\left|f_m\right|^2
 e^{im\trel \gamma}
 \end{eqnarray}
 \end{subequations}
no matter which Fourier transform is chosen. The averaged spectral function as defined in Eq. \eqref{eqn:averaged_Spectralfunction} yields
 \begin{subequations}
 \begin{eqnarray}
\bar{A}\left(\omega, \vect{k}\right)
=
-\frac{1}{\pi}\mathrm{Im}
\left[
-\frac{i}{\hbar}\sum_m\left|f_m\right|^2
\right.
\\
\times
\left.
\lim_{\eta\rightarrow 0^+}
\int_0^\infty 
e^{i\trel\left(\omega+m\gamma-\frac{\varepsilon\left(\vect{k}\right)}{\hbar}J_0\left(\frac{E_0}{\hbar}\right)+i\eta\right)
}\mathrm{d}\,\trel
\right]
\\
=
\sum_m
\frac{\left|f_m\right|^2}{\hbar}
\delta\left(
\omega+m\gamma-\frac{\varepsilon\left(\vect{k}\right)}{\hbar}J_0\left(\frac{E_0}{\hbar}\right)\right)\,.
 \end{eqnarray}
 \end{subequations}
This is manifestly non negative and completes the proof.


\begin{thebibliography}{44}

\bibitem{Floquet1883}
G. Floquet, Ann. de l'Ecole Norm. Sup. \textbf{12}  47-88 (1883)

\bibitem{Lindner2011}
N.H. Lindner, G. Refael, V. Galitski,
Nat. Phys. \textbf{7} 490-495 (2011)

\bibitem{Kitagawa2010}
T. Kitagawa, E. Berg, M. Rudner, E. Demler,
Phys. Rev. B
\textbf{82} 235114 (2010)

\bibitem{Gedik2013}
Y. H. Wang, H. Steinberg, P. Jarillo-Herrero, N. Gedik,
Science \textbf{342} 453
(2013)

\bibitem{Fregoso2013}
B. M. Fregoso, Y. H. Wang, N. Gedik, V. Galitski,
Phys. Rev. B \textbf{88} 155129 (2013)
\bibitem{Sentef2015}
M.A. Sentef, M. Claassen, A.F. Kemper, B. Moritz, T. Oka, J.K. Freericks, T.P. Devereaux,
Nat. Commun. \textbf{6} 7047 (2015)
\bibitem{Claassen2016}
M. Claassen, C. Jia, B. Moritz, T.P. Devereaux,
Nat. Commun. \textbf{7} 13074 (2016)
\bibitem{Tang2017}
S. Tang, C. Zhang, D. Wong, Z. Pedramrazi, H.-Z. Tsai, C. Jia,
B. Moritz, M. Claassen, H. Ryu, S. Kahn
\textit{et al.},
Nat. Phys.
\textbf{13}
683
(2017)
\bibitem{Peierls1933}
R. Peierls,
Z. Phys. \textbf{80} 763-791.
(1933)
\bibitem{Aoki2014}
H. Aoki, N. Tsuji, M. Eckstein, M. Kollar, T. Oka, P. Werner,
Rev. Mod. Phys. \textbf{86} 779 (2014)
\bibitem{Freericks2006_3}
J. K. Freericks, V. M. Turkowski, V. Zlati\'{c},
Phys. Rev. Lett. \textbf{97} 266408 (2006)


\bibitem{Slater1954} J. C. Slater, G.F. Koster, Phys. Rev. \textbf{94}, 1408 (1954)

\bibitem{Metzner1989} W. Metzner, D. Vollhardt, Phys. Rev. Lett. \textbf{62}, 324 (1989)

\bibitem{Jackson2001}
J. D. Jackson, L. B. Okun, Rev. Mod. Phys. \textbf{73} 663 (2001)

\bibitem{Jauho1984} A. P. Jauho, J. W. Wilkins, Phys. Rev. B, \textbf{29},1919 (1984)

\bibitem{Turkowski2005} V. Turkowski, J.K. Freericks, Phys. Rev. B. \textbf{71} 085104 (2005)
\bibitem{Gruber2001}
Ch. Gruber, N. Macris, Ph. Royer, J. K. Freericks,
Phys. Rev. B. 
\textbf{63} 165111 (2001)
\bibitem{Freericks2009_1}
 J. K. Freericks, H. R. Krishnamurthy, Th. Pruschke, 
 Phys. Rev. Lett. \textbf{102} 136401
 (2009)
\bibitem{Freericks2016_2}
J. K. Freericks, H. R. Krishnamurthy
Photonics \textbf{3} 58 (2016)
\bibitem{Wigner1932} E. P. Wigner, Phys. Rev. \textbf{40}, 749 (1932)


\bibitem{Grifoni1998}
M. Grifoni, P. H\"anggi, Phys. Rep. \textbf{304} (1998) 229-354

\bibitem{supp_material}
See Supplemental Material at \url{http://link.aps.org/supplemental/10.1103/PhysRevB.98.035138}
for the raw data for all figures in
the paper.

\bibitem{Kennes2018}
 D. M. Kennes, A. de la Torre, A. Ron, D. Hsieh, A. J. Millis, 
 Phys. Rev. Lett. \textbf{120} 127601
 (2018)



\end{thebibliography}
\end{document}